\documentclass[letter,twocolumn]{jpsj3}

\usepackage{txfonts}

\usepackage{amsmath}
\usepackage{amsfonts}
\usepackage{bm}
\usepackage{graphicx}
\usepackage{color}

\title{Solving the Bose-Hubbard model with machine learning
}

\author{Hiroki Saito}
\inst{Department of Engineering Science, University of
Electro-Communications, Tokyo 182-8585, Japan}

\abst{Motivated by the recent successful application of artificial neural
networks to quantum many-body problems [G. Carleo and M. Troyer, Science
{\bf 355}, 602 (2017)], a method to calculate the ground state of the
Bose-Hubbard model using a feedforward neural network is proposed.
The results are in good agreement with those obtained by exact
diagonalization and the Gutzwiller approximation.
The method of neural-network quantum states is promising for solving
quantum many-body problems of ultracold atoms in optical lattices.
}

\begin{document}
\maketitle

Machine learning with artificial neural networks has attracted a great
deal of public attention following the defeat of professional Go players
by the AlphaGo computer program~\cite{AlphaGo}.
Machine learning techniques are growing dramatically and are being applied
to wide areas in engineering and science.
In physics, neural-network techniques have been applied to various
problems~\cite{Ohtsuki,Carra,Nieu,Tanaka,Zhang,Ovch,Torlai,Huang}, such as
identification of phase
transitions~\cite{Ohtsuki,Carra,Nieu,Tanaka,Zhang}.

Recently, it was shown that artificial neural networks can be used to
solve quantum many-body problems~\cite{Carleo}.
The main difficulty in solving quantum many-body problems through
numerical calculations is that the Hilbert space exponentially diverges as
the number of particles increases, and the large amount of data needed to
express the wave functions exceeds the capacity of computers.
In Ref.~\citen{Carleo}, it was proposed that information of the wave
function be stored in the neural network, and when a basis (e.g., a spin
configuration $\uparrow\downarrow \cdots$) is input to the network, the
corresponding expansion coefficient (probability amplitude) of the wave
function is obtained as an output.
Neural networks can recognize and extract features from large amounts of
data.
For example, in image recognition, features of image data (not image data
themselves) are stored in the network.
In a similar manner, we expect that features of wave functions can be
extracted and efficiently stored in a neural network.
In Ref.~\citen{Carleo}, it was demonstrated that the restricted Boltzmann
machine can solve quantum many-body spin problems (transverse-field
Ising model and antiferromagnetic Heisenberg model) very efficiently.
The entanglement properties of such neural-network states were
investigated in Ref.~\citen{Deng}.
The expressibility of artificial neural networks of many-body wave
functions was investigated in Ref.~\citen{Cai}.

\begin{figure}
\begin{center}
\includegraphics[width=8.5cm]{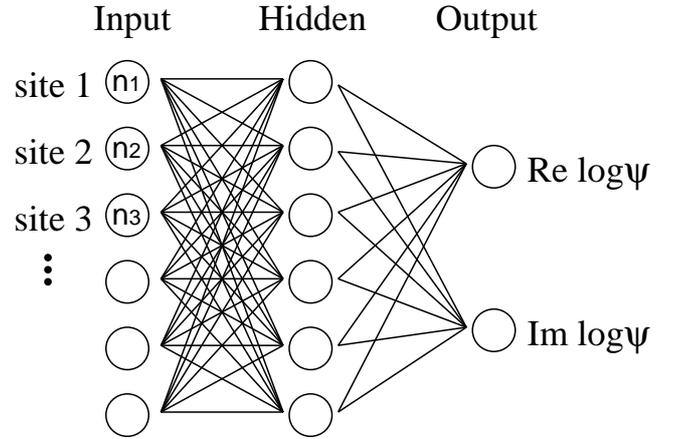}
\end{center}
\caption{
Schematic diagram of the artificial neural network used to solve the
Bose-Hubbard model.
The number of particles at each site is assigned to the input layer, and
the corresponding value of the wave function is obtained from the output
layer.
The units in the adjacent layers are fully connected.
}
\label{f:schematic}
\end{figure}
In the present Letter, the method in Ref.~\citen{Carleo} is extended to
treat many bosons on a lattice, i.e., the Bose-Hubbard model.
Instead of the restricted Boltzmann machine used in
Ref.~\citen{Carleo}, a fully-connected feedforward network, as shown in
Fig.~\ref{f:schematic}, is used.
The quantum state is expanded by the Fock states $|\Psi\rangle = \sum
\psi(n_1, n_2, \cdots) |n_1, n_2, \cdots \rangle = \sum \psi(\bm{n})
|\bm{n} \rangle$, where $n_i$ is the
number of particles at the $i$th site.
When a set of integers $\bm{n}$ is input to the network, the value of the
wave function $\psi(\bm{n})$ is obtained from the output layer.
We attempt to optimize the parameters of the network so that the output
$\psi(\bm{n})$ is close to the ground-state wave function.

The Bose-Hubbard Hamiltonian is given by
\begin{equation} \label{H}
\hat H = -J \sum_{\langle i j \rangle} \hat a_i \hat a_j^\dagger
+ \sum_i \left[ V_i \hat n_i + \frac{U}{2} \hat n_i (\hat n_i - 1)
\right],
\end{equation}
where $J$ is the tunneling coefficient, $\sum_{\langle i j \rangle}$
denotes the sum over all pairs of adjacent sites, $V_i$ is the
site-dependent potential, $\hat n_i = \hat a_i^\dagger \hat a_i$ is
the number operator, and $U$ is the on-site interaction energy.
The system exhibits superfluidity for $U / J \lesssim 1$ and enters the
Mott insulator state for $U / J \gg 1$.
The number of the Fock-state bases $|\bm{n} \rangle$ is $(N + M - 1)! /
[N! (M - 1)!] \equiv N_B$ and exponentially increases with $N$ and $M$,
where $N = \sum_i n_i$ is the number of particles, and $M$ is the number
of sites.

The feedforward network in Fig.~\ref{f:schematic} operates as follows.
The integers $\bm{n}$ are set to the input units as $u^{(0)}_j = n_j$,
where the number of units in the input layer is $M$.
The values of the hidden units are calculated as
\begin{equation} \label{ff1}
u^{(1)}_k(\bm{n}) = \sum_{j=1}^{M} W^{(1)}_{kj} n_j + h^{(1)}_k.
\end{equation}
We adopt the hyperbolic tangent as an activation function, and the output
units become
\begin{equation} \label{ff2}
u^{(2)}_m(\bm{n}) = \sum_{k=1}^{N_H} W^{(2)}_{mk} \tanh u^{(1)}_k(\bm{n})
+ h^{(2)}_m,
\end{equation}
where $N_H$ is the number of units in the hidden layer, and $m = 1, 2$.
The weights $W^{(1)}_{kj}$ and $W^{(2)}_{mk}$ and the biases $h^{(1)}_k$
and $h^{(2)}_m$ are real.
The wave function is thus given by
\begin{equation} \label{psi}
\psi(\bm{n}) = \exp[u^{(2)}_1(\bm{n}) + i u^{(2)}_2(\bm{n})].
\end{equation}

An expectation value of a quantity $\hat A$,
\begin{equation} \label{expect}
\langle \hat A \rangle = \frac{\sum_{\bm{n}, \bm{n}'} \psi^*(\bm{n})
\langle \bm{n} | \hat A | \bm{n}' \rangle \psi(\bm{n}')}{\sum_{\bm{n}}
|\psi(\bm{n})|^2},
\end{equation}
is calculated by the Monte Carlo method with Metropolis sampling.
Given $\bm{n}_1$ and $\bm{n}_2$, the probability that $\bm{n}_1
\rightarrow \bm{n}_2$ is adopted, $|\psi(\bm{n}_2) / \psi(\bm{n}_1)|^2$,
can be calculated from the network, and we can then sample $\bm{n}$ with
probability $|\psi(\bm{n})|^2 / \sum_{\bm{n}'} |\psi(\bm{n}')|^2$.
The expectation value in Eq.~(\ref{expect}) is therefore stochastically
calculated as
\begin{equation} \label{expect2}
\left\langle \sum_{\bm{n}'} \langle \bm{n} | \hat A | \bm{n}' \rangle
\frac{\psi(\bm{n}')}{\psi(\bm{n})} \right\rangle_M
\equiv \left\langle \tilde A \right\rangle_M,
\end{equation}
where $\langle \cdots \rangle_M$ denotes the average over the Metropolis
sampling of $\bm{n}$.
When the matrix $\langle \bm{n} | \hat A | \bm{n}' \rangle$ is sparse, the
sum over $\bm{n}'$ in Eq.~(\ref{expect2}) can easily be calculated.

The network parameters $\bm{W}$ and $\bm{h}$ in Eqs.~(\ref{ff1}) and
(\ref{ff2}) are optimized so that the expectation
value of the Hamiltonian $\langle \hat H \rangle$ becomes minimum.
Although the stochastic reconfiguration method~\cite{Sorella} is more
stable~\cite{Carleo}, for simplicity, we use the steepest descent method
for the optimization.
The derivative of the energy with respect to the network parameter is
given by
\begin{eqnarray} \label{grad}
\frac{\partial \langle \hat H \rangle}{\partial w}
& = & 2 {\rm Re} \Biggl[ \frac{\sum_{\bm{n}, \bm{n}'} O_w^*(\bm{n})
\psi^*(\bm{n}) \langle \bm{n} | \hat H | \bm{n}' \rangle \psi(\bm{n}')}
{\sum_{\bm{n}} |\psi(\bm{n})|^2} 
\nonumber \\
& & - \langle \hat H \rangle
\frac{\sum_{\bm{n}} O_w^*(\bm{n}) |\psi(\bm{n})|^2}
{\sum_{\bm{n}} |\psi(\bm{n})|^2} \Biggr]
\nonumber \\
& \simeq & 2 {\rm Re} \left( \langle O_w^* \tilde H \rangle_M
- \langle O_w^* \rangle_M \langle \tilde H \rangle_M \right),
\end{eqnarray}
where $w$ is one of the network parameters $\bm{W}$ or $\bm{h}$, and
\begin{equation} \label{O}
O_w(\bm{n}) = \frac{1}{\psi(\bm{n})}
\frac{\partial\psi(\bm{n})}{\partial w}.
\end{equation}
The derivative in Eq.~(\ref{O}) is calculated using Eqs.~(\ref{ff1}),
(\ref{ff2}), and (\ref{psi}).
The network parameters are updated as
\begin{equation} \label{update}
w \rightarrow w - \gamma
\frac{\partial \langle \tilde H \rangle_M}{\partial w},
\end{equation}
where $\gamma$ is a rate controlling the parameter change.
The value of $\gamma$ is taken to be $10^{-1}$-$10^{-3}$.
Typically, $10^3$-$10^4$ updates are needed for sufficient convergence.
The average $\langle \cdots \rangle_M$ in each update step is
calculated from $10^3$ samples, and the final energy is calculated from
$10^4$ samples.
The network parameters are initialized by random numbers with a normal
distribution, where the standard deviation is taken to be $\sim 0.1$.

\begin{figure}
\begin{center}
\includegraphics[width=8.5cm]{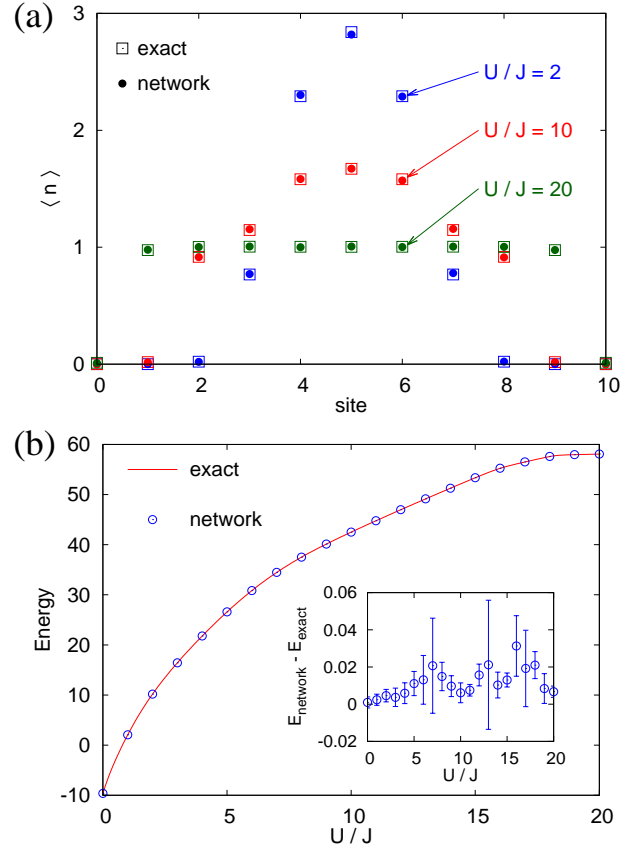}
\end{center}
\caption{
(Color online) Ground state of the one-dimensional Bose-Hubbard model for
$M = 11$ sites and $N = 9$ particles with a harmonic confinement in
Eq.~(\ref{V}).
The numbers of hidden units is $N_H = 20$.
(a) Distribution of particle numbers for $U / J = 2$, 10, and 20.
The circles and squares are obtained by the present method and
exact diagonalization, respectively.
(b) The ground-state energy as a function of $U / J$ obtained by the
present method (circles) and exact diagonalization (line).
The inset shows the difference between them, where the error bars
represent the statistical error calculated from ten values.
}
\label{f:1d}
\end{figure}
First, we consider a one-dimensional system with $M = 11$ sites and $N =
9$ particles.
In experiments of ultracold atoms in an optical lattice, a weak harmonic
potential is superimposed over the lattice potential due to the profile of
laser beams~\cite{Bloch}, and we take the site-dependent potential
as~\cite{Note}
\begin{equation} \label{V}
V_j = V (j - 5)^2 \qquad (j = 0, 1, \cdots, 10),
\end{equation}
where we take $V = J$ in the following calculations.
The number of hidden units is taken to be $N_H = 20$.
Figure~\ref{f:1d}(a) shows the expectation value of particle numbers at
each site.
As $U / J$ is increased, the particle distribution expands and the Mott
insulator state is reached for $U / J = 20$.
Figure~\ref{f:1d}(b) shows the ground-state energy as a function of $U /
J$.
In Fig.~\ref{f:1d}, the results obtained by exact diagonalization of the
Hamiltonian using the Lanczos method are also shown.
We find that the results obtained by the neural-network method are in
excellent agreement with the exact results.
The overlap,
\begin{equation}
\frac{|\sum_{\bm{n}} \psi^*(\bm{n}) \psi_{\rm exact}(\bm{n})|^2}
{\sum_{\bm{n}} |\psi(\bm{n})|^2},
\end{equation}
between the wave function stored in the network $\psi(\bm{n})$ and the
normalized exact wave function $\psi_{\rm exact}(\bm{n})$ is larger than
0.99.
Note that the number of network parameters, $\bm{W}$ and $\bm{h}$, is
$(11 + 1) \times 20 + (20 + 1) \times 2 = 282$, whereas the number of
bases for the exact diagonalization is $N_B = 125970$, which indicates
that the information of the wave function is efficiently stored in the
neural network.

\begin{figure}
\begin{center}
\includegraphics[width=8.5cm]{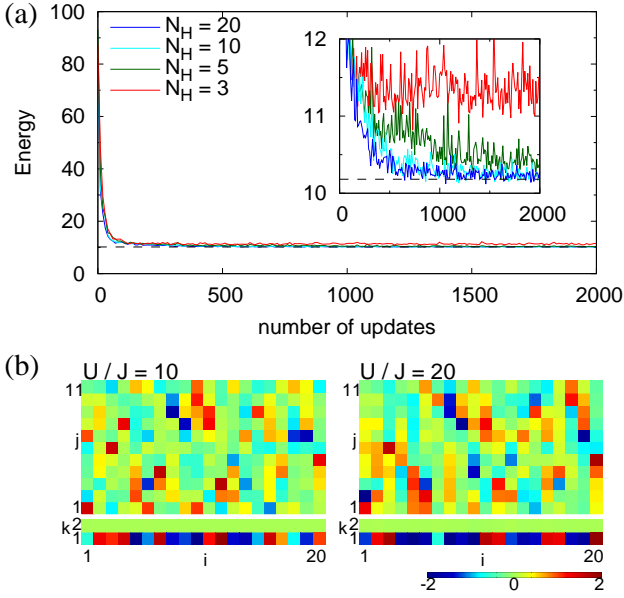}
\end{center}
\caption{
(Color online) (a) Energy of the system as a function of optimization
steps for $U / J = 2$.
The numbers of hidden units are $N_H = 20$, 10, 5, and 2.
Other conditions are the same as those in Fig.~\ref{f:1d}.
The dashed line indicates the exact energy of the ground state.
The inset shows a magnified section of the main panel. 
(b) The network parameters $W^{(1)}_{ij}$ and $W^{(2)}_{ik}$ optimized in
Fig.~\ref{f:1d}.
}
\label{f:depend}
\end{figure}
Figure~\ref{f:depend}(a) shows the optimization process of the neural
network.
The energy first decreases quickly as the network parameter is updated as
in Eq.~(\ref{update}), and then gradually converges to the final value.
As the number of hidden units $N_H$ is decreased, the final value of the
energy deviates from the correct value, because the ability to represent
the quantum state decreases as the number of network parameters decreases.
Nevertheless, we can obtain the qualitative properties of quantum
many-body systems, even for rather small $N_H$.
To see the internal state of the network, the values of
$\bm{W}$ after the optimization are shown in Fig.~\ref{f:depend}(b).
However, it is difficult to capture the features of the network.

In Fig.~\ref{f:depend}(a), the computational time is several seconds for
$N_H = 20$ using my work station (Intel Xeon E5-2697A v4), which is
comparable to or shorter than the computational time for the exact
diagonalization of the same problem using the ARPACK library.
The computational amount for the exact diagonalization is $O(N_B)$ and
exponentially increases with $N$ and $M$, while the computational amount
for the present method is $O(M N_H) \times$ number of updates.

\begin{figure}
\begin{center}
\includegraphics[width=8.5cm]{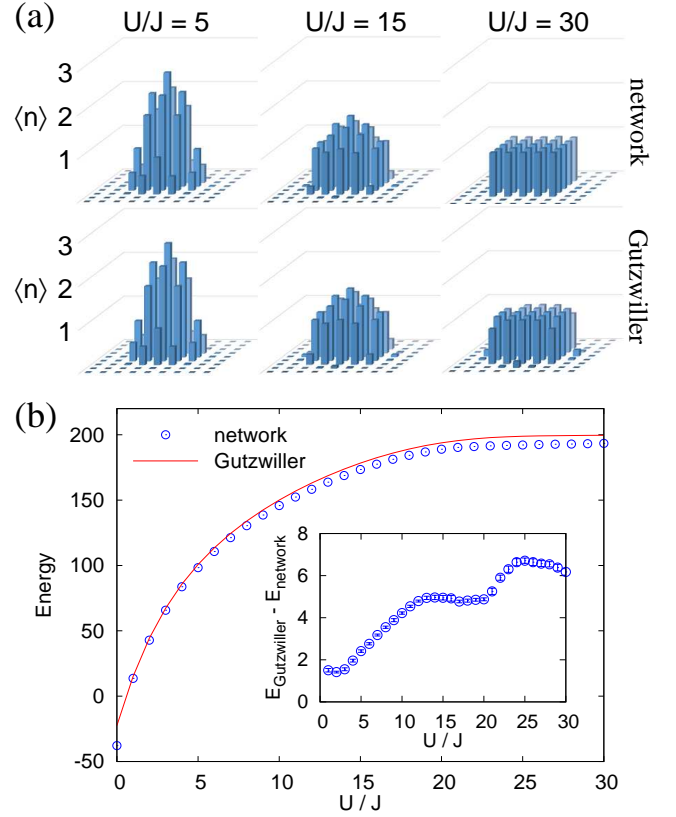}
\end{center}
\caption{
(Color online) Ground state of the two-dimensional Bose-Hubbard model for
$M = 9 \times 9$ sites and $N = 25$ particles with a harmonic confinement
in Eq.~(\ref{V2}).
The numbers of hidden units is $N_H = 40$.
(a) Distribution of particle numbers for $U / J = 5$, 15, and 30.
The upper and lower panels are obtained by the present method and by the
Gutzwiller approximation, respectively.
(b) The ground-state energy as a function of $U / J$ obtained by the
present method (circles) and the Gutzwiller approximation (line).
The inset shows the difference between them, where the error bars
represent the statistical error calculated from ten values.
}
\label{f:2d}
\end{figure}
Next, we consider a two-dimensional system with $N = 25$ particles at
$M = 9 \times 9$ sites.
The site-dependent potential has the form
\begin{equation} \label{V2}
V_{j_x, j_y} = V [(j_x - 4)^2 + (j_y - 4)^2] \qquad (j_x, j_y = 0, 1,
\cdots 8),
\end{equation}
where we take $V = 2 J$.
Figure~\ref{f:2d}(a) shows the expectation value of the particle number
distribution.
The upper panels are obtained by the present neural-network method, and
the lower panels are obtained by the Gutzwiller
approximation~\cite{Jaksch}.
The distributions obtained by the two methods agree well, except at the
edge of the Mott insulator state for $U / J = 30$.
Figure~\ref{f:2d}(b) shows the ground-state energy as a function of $U /
J$ obtained by the neural-network method and the Gutzwiller
approximation.
smaller than that by the Gutzwiller approximation, which implies that the
former is better than the latter.

In the above calculations, the network parameters were optimized by the
stochastic gradient method in Eq.~(\ref{update}).
However, the optimization was found to sometimes be trapped by a local
minimum of the energy.
A simple manner to avoid the local minima is to change the parameter
adiabatically.
For example, if we start from $U = 0$ and slowly increase $U$ during the
optimization procedure, the ground state is maintained according to the
adiabatic theorem.
Figures~\ref{f:1d}(b) and \ref{f:2d}(b) were obtained in this manner.
In the field of deep learning (machine learning with multi-layer neural
networks), various techniques to circumvent undesired states of networks
in optimization processes have been developed, such as
dropout~\cite{Sriv}, which may also be efficient for avoiding local minima
in quantum many-body problems.

In conclusion, a method to obtain the ground state of the Bose-Hubbard
model using an artificial neural network was proposed.
It was demonstrated that the approximate ground state can be obtained by a
simple optimization scheme of the network parameters.
The results for one-dimensional and two-dimensional systems are in good
agreement with those obtained by exact diagonalization and by the
Gutzwiller approximation, even for small networks, which implies that the
information of many-body quantum states is efficiently stored in the
artificial neural networks.
There may be a variety of extensions of the present study.
The present method can easily be extended to multiple layers, which is
interesting from the viewpoint of deep learning.
Much larger systems may be explored using an existing neural-network
framework optimized for GPU computing, which enables us to obtain phase
structures in the thermodynamic limit.
Fermions can be treated in a similar manner.
Atoms with spin degrees of freedom on a lattice is also an area of
interest.
Time evolution can be implemented using the method in
Ref.~\citen{Carleo}.
Extension of the discrete lattice to continuous space will be a
challenging task.

\begin{acknowledgments}
This work was supported by JSPS KAKENHI Grant Numbers JP16K05505,
JP17K05595, JP17K05596, and JP25103007.
\end{acknowledgments}

\end{document}